\documentclass{ws-ijmpd}

\begin{document}

\markboth{Das, Ray, Radinschi, Rahaman and Ray}
{static charged fluid spheres}

\catchline{}{}{}{}{}
%
%%%%%%%%%%%%%%%%%%%%%%%%%%%%%%%%%%%%%%%%%%%%%%%%%%%%%%%%%%%%%%%%%%%%

\title{Isotropic cases of static charged fluid spheres in general relativity}

\author{Basanti Das}
\address{Belda Prabhati Balika Vidyapith, Belda,
West Midnapur 721 424, W.B., India} 

\author{Pratap Chandra Ray}
\address{Department of Mathematics, Government College of Engineering and
Leather Technology, Kolkata 700 098, West Bengal, India \\raypratap1@yahoo.co.in} 

\author{Irina Radinschi}
\address{Department of Physics, ``Gh. Asachi'' Technical University, Iasi,
700050, Romania \\radinschi@yahoo.com}

\author{Farook  Rahaman} 
\address{Department of Mathematics, Jadavpur University, Kolkata 700
032, West Bengal, India \\farook\_rahaman@yahoo.com}

\author{Saibal Ray}
\address{Department of Physics, Government College of
Engineering \& Ceramic Technology, Kolkata 700 010, West Bengal,
India\\ saibal@iucaa.ernet.in}

\maketitle

\begin{history}
\received{Day Month Year}
\revised{Day Month Year}
\comby{Managing Editor}
\end{history}

\begin{abstract}
In this paper we study the isotropic cases of static charged fluid
spheres in general relativity. For this purpose we consider two
different specialization and under these we solve the
Einstein-Maxwell field equations in isotropic coordinates. The
analytical solutions thus we obtained are matched to the exterior
Reissner-Nordstr\"{o}m solutions which concern with the values for
the metric coefficients $e^{\nu }$ and $e^{\mu }$. We derive the
pressure, density, pressure-to-density ratio at the centre of the
charged fluid sphere and boundary $R$ of the star. Our conclusion
is that static charged fluid spheres provide a good connection to
compact stars.
\end{abstract}


\section{Introduction}
~~~~The topic of static charged fluid spheres in general
relativity is a challenging issue and has given rise to
interesting studies, even if their number is not so high.
Considering the research of the interior of stars in connection to
their late stage evolution when the general relativistic effects
play an important role, we notice the interesting work of Tolman
\cite{Tolman1939} that yields a class of solutions for the static,
spherically symmetric equilibrium fluid distribution. This
important result was followed by some generalizations made by
Wyman \cite{Wyman1949}, Leibovitz \cite{Leibovitz1969} and Whitman
\cite{Whitman1977}. After them, Bayin \cite{Bayin1978} used the
idea of the method of quadratures and gave new astrophysical
solutions for the static fluid spheres.

In recent years, Ray and Das \cite{Ray2002,Ray2004,Ray2007a} and
Ray et. al. \cite{Ray2007b} have developed some interesting
solutions for the static charged fluid spheres in general
relativity following the above line of thinking. Ray and Das
\cite{Ray2002} performed the charged generalization of Bayin's
work \cite{Bayin1978} related to Tolman's type astrophysically
interesting aspects of stellar structure \cite{Tolman1939}. In
this light, they actually considered that even in the case of
stellar astrophysics there are physical implications of the
Einstein-Maxwell field equations. In other papers Ray and Das
\cite{Ray2004,Ray2007a} performed a study of some previous
solutions considering the phenomenological connection of the
gravitational field to the electromagnetic field, and demonstrated
the purely electromagnetic origin of the charged relativistic
stars given by Tolman \cite{Tolman1939} and Bayin
\cite{Bayin1978}. The existence of this type of astrophysical
solutions is thought to be a probable extension of Lorentz's
conjecture \cite{Lorentz1904} that electron-like extended charged
particle possesses only `electromagnetic mass' and no `material
mass'. In their latest work Ray et al. \cite{Ray2007b} have
considered Tolman-Bayin type static charged fluid spheres in
general relativity and they have found out many interesting
results, specially the cases which give support to the charged
spherical models in connection to normal stars. It is argued by
them \cite{Ray2007b} that due to the inclusion of charge and by a
suitable choice of charge part, pressure and density function
could be a decreasing function of radius from centre to surface in
contrary to Bayin's case \cite{Bayin1978}.

In this connection we also notice other two works, the first one
elaborated by Varela \cite{Varela2007}, which presents a neutral
perfect fluid core bounded by a charged thin shell, and which
demonstrates that it is possible to construct extended
Reissner-Nordstr\"{o}m sources with everywhere positive mass
density, classical electron radius, electromagnetic mass, and
everywhere non-negative gravitational mass. The other work was
done by Ivanov \cite{Ivanov2002} that studied the interior perfect
fluid solutions for the Reissner-Nordstr\"{o}m metric using a new
classification scheme. In addition, he found general formulae in
some particular cases, presented explicit new global solutions and
made a revision of some known solutions. In connection to the
singularity problem it is argued by Ivanov \cite{Ivanov2002} that
the presence of the charge function serves as a safety valve,
which absorbs much of the fine-tuning necessary in the uncharged
case. However, it is also believed by some authors
\cite{Felice1995,Sharma2001} that in the presence of charge the
gravitational collapse of a spherically symmetric distribution of
matter to a point singularity may be avoided through
counterbalancing of the gravitational attraction by the repulsive
Coulombian force in addition to the thermal pressure gradient due
to fluid.

In continuation of the above theme, especially that of Ray et al.
\cite{Ray2007b}, in the present article we have tried to solve the
Einstein-Maxwell field equations in isotropic coordinate system
and derived expressions for pressure and density. We have found
out conditions for the boundary of the charged sphere. The
exterior Reissner-Nordstr{\"{o}}m solution is compared  and
constants of integrations are expressed in terms of mass and
radius. The mass-radius and mass-charge relations have been found
out for various cases of the charged matter distribution along
with the pressure-to-density ratio at the centre of the charged
sphere.

Our paper is organized as follows: in Section 2 we introduce the
Einstein-Maxwell field equations for the static charged fluid
spheres in general relativity, we determine their analytical
solutions making some assumption for the metric coefficients
$e^{\nu}$ and $e^{\mu}$ and we consider two particular cases. In
Section 3 we find out boundary conditions, which consist in a
vanishing value for the pressure and a specific value for the
boundary $R$ of the star. The expression for the radius of the
star $R$ is established in Section 4 that is dedicated to a
detailed analysis of the obtained solutions. The role of some
parameters and integration constants in the evolution of pressure
$p$, density $\rho$ and radius of the star $R$ is discussed, with
emphasis on some particular values. In Conclusions we enlighten
the physical significance of the results.\\

\section{The Einstein-Maxwell field equations and their analytical solutions}
~~~~The static spherically symmetric matter distribution
corresponding to the isotropic line element is given by
\begin{eqnarray}
ds^{2}=e^{\nu }dt^{2}-e^{\mu }(dr^{2}+r^{2}d\theta
^{2}+r^{2}sin^{2}\theta d\phi ^{2}),  \label{metric}
\end{eqnarray}
where $\nu $ and $\mu $ are functions of the radial coordinate $r$
only.

The Einstein-Maxwell field equations for the above line element
can be written as
\begin{eqnarray}
e^{-\mu }\left( \frac{{\mu ^{\prime }}^{2}}{4}+\frac{\mu ^{\prime
}\nu
^{\prime }}{2}+\frac{\mu ^{\prime }+\nu ^{\prime }}{r}\right) +\frac{q^{2}}{%
r^{4}}=8\pi p,  \label{mu1}
\end{eqnarray}
\begin{eqnarray}
e^{-\mu }\left( \frac{\mu ^{{\prime }{\prime }}}{2}+\frac{\nu ^{{\prime }{%
\prime }}}{2}+\frac{{\nu ^{\prime }}^{2}}{4}+\frac{\mu ^{\prime
}+\nu ^{\prime }}{2r}\right) -\frac{q^{2}}{r^{4}}=8\pi p,
\label{mu2}
\end{eqnarray}
\begin{eqnarray}
-e^{-\mu }\left( \mu ^{{\prime }{\prime }}+\frac{{\mu ^{\prime }}^{2}}{4}+%
\frac{2\mu ^{\prime }}{r}\right) -\frac{q^{2}}{r^{4}}=8\pi \rho ,
\label{mu3}
\end{eqnarray}
where the total charge with a sphere of radius $r$ in terms of the $4$%
-currents $J^{i}$ is
\begin{eqnarray}
q(r)=r^{2}E(r)=4\pi \int_{0}^{r}J^{0}r^{2}e^{(\mu +\nu )/2}dr,
\end{eqnarray}
$E$ being the intensity of the electric field. Here $p$ and $\rho
$ are the pressure and matter-energy density, respectively. We
have used prime to denote derivative with respect to radial
coordinate $r$ only. The field equations without the charge $q$
have been studied by Buchdahl \cite{Buchdahl1966} and Bayin
\cite{Bayin1978} who found physically meaningful solutions.

Equating equations (\ref{mu1}) and (\ref{mu2}) we get a
differential equation by assuming $e^{\nu}=A\Phi^{-a}$ and
$e^{\mu}=B\Phi^{b}$ in the form
\begin{eqnarray}
\Phi^{{\prime}{\prime}} - c \frac{{\Phi^{\prime}}^{2}}{\Phi} - \frac {%
\Phi^{\prime}}{r}= \frac{4B}{b-a} \Phi^{b+1}\frac{q^{2}}{r^{4}},
\label{phi1}
\end{eqnarray}
where $\Phi=\Phi(r)$ and $c=(\frac{1}{2}b^{2} - \frac{1}{2}a^{2}
-ab + b -a)/(b-a)$.

Hence, to solve the above equation (\ref{phi1}) for $\Phi(r)$ we
make use of the \textit{ansatz} $q(r)^{2} = K^{2} \Phi(r)^{d}$ so
that the above equation becomes
\begin{eqnarray}
\Phi^{{\prime}{\prime}} - c \frac{{\Phi^{\prime}}^{2}}{\Phi} - 
\frac{\Phi^{\prime}}{r}= \frac{4B}{b-a} \Phi^{c} \frac{K^{2}}{r^{4}},
\end{eqnarray}
with $c=b+1+d$ where $d$ behaves as the polynomial index. A direct
solution for the above second order differential equation is
difficult to find out. However, we can set different conditions
for the parameters to get a solvable equation. Let us therefore
consider the following two cases.\\

\subsection{The case for $c = 1$ and $d=-b$}
~~~~For the specification $c = 1$, when $d=-b$, we get the
following solutions
\begin{equation}
\Phi=C_1 e^{[\frac{B}{2(b-a)} \frac{K^2}{r^2} + C_0 r^2]},
\end{equation}
\begin{equation}
e^{\nu}=A {C_1}^{-a} e^{-a[\frac{B}{2(b-a)} \frac{K^{2}}{r^{2}} +
C_{0} r^{2}]},
\end{equation}
\begin{equation}
e^{\mu}=B {C_1}^{b} e^{b[\frac{B}{2(b-a)} \frac{K^{2}}{r^{2}} +
C_{0} r^{2}]}.
\end{equation}

\subsection{The case for $c \neq1$ and $d=-\frac{1}{2}(b^{2}+a^{2})/(b-a)$}
~~~~For this choice of $c$ and $d$ the solutions set becomes
\begin{equation}
\Phi ^{1-c}=\frac{B(1-c)}{2(b-a)}\frac{K^{2}}{r^{2}}+C_{0}r^{2}
+C_{1},
\end{equation}
\begin{equation}
e^{\nu }=A
\left[\frac{B(1-c)}{2(b-a)}\frac{K^{2}}{r^{2}}+C_{0}r^{2}
+C_{1}\right]^{-a/(1-c)},
\end{equation}
\begin{equation}
e^{\mu
}=B\left[\frac{B(1-c)}{2(b-a)}\frac{K^{2}}{r^{2}}+C_{0}r^{2}
+C_{1}\right]^{b/(1-c)}.
\end{equation}

We notice from the above two subcases that $C_{0}$ and $C_{1}$
represent integration constants and their expressions will be
determined in Section $4$ in terms of $m$ and $R$.\\

\section{Boundary conditions}
~~~~The exterior field of a spherically symmetric static charged
fluid distribution described by the metric (1) in isotropic
coordinates is the unique Reissner-Nordstr{\"{o}}m solution
\begin{eqnarray}
ds^{2}=\left( 1-\frac{2m}{r}+\frac{q^{2}}{r^{2}}\right)
dt^{2}-\left( 1- \frac{2m}{r}+\frac{q^{2}}{r^{2}}\right)
^{-1}dr^{2} -r^{2}(d\theta ^{2}+sin^{2}\theta d\phi ^{2}),
\end{eqnarray}
which by the radial coordinate transformation \cite{Vogt2004}
\begin{equation}
r=r^{\prime}\left( 1+\frac{m}{2r^{\prime}}+\frac{q}{2r^{\prime}}\right) \left(
1+\frac{m}{2r^{\prime}}-\frac{q}{2r^{\prime}}\right) ,
\end{equation}
takes the form
\begin{equation}
ds^2=\frac{\left(1-\frac{m^2}{4{r^{\prime}}^2}+\frac{q^2}{4{r^{\prime}}^2}\right)^2}{\left[\left(
1+\frac{m}{2r^{\prime}}\right)^2 -\frac{q^2}{4{r^{\prime}}^2}\right]^2} dt^2-
\left[\left(1+\frac{m}{2r^{\prime}}\right)^2-\frac{q^2}{4{r^{\prime}}^2}\right]^2 (dr^2+r^2 d \theta
^2+r^2sin^2\theta d\phi^2).
\end{equation}

Therefore, matching at $r^{\prime}=R$ we get
\begin{equation}
e^{\nu(R)} = \frac{\left(1-\frac{m^2}{4R^2}+\frac{q^2}{4R^2}\right)^2}{\left[\left(
1+\frac{m}{2R}\right)^2 -\frac{q^2}{4R^2}\right]^2} ,
\end{equation}
\begin{eqnarray}
e^{\mu(R)} = \left[\left(1+\frac{m}{2R}\right)^2-\frac{q^2}{4R^2}\right]^2,
\end{eqnarray}
where $R$ is the boundary of the star. At the boundary pressure is
zero. Using this we will get an expression for $R$ in the next
section.\\

\section{The study of the solutions}

\subsection{The case for $c = 1$ and $d=-b$}
\begin{eqnarray}
8\pi p & =\frac{{C_1}^{-b}}{B}[\frac{b(b-2a)}{4}(\frac{-B} {b-a}
\frac{K^2}{r^3}+2C_0 r)^2 + (\frac{b-a}{r})(\frac{-B}{b-a}\frac
{K^2}{r^3}+2C_{0}r)} +\frac{BK^2}{r^4}]e^{-b\widetilde{D(r)},
\end{eqnarray}
\begin{eqnarray}
8\pi\rho &
=-\frac{{C_1}^{-b}}{B}[\frac{BbK^2}{b-a}\frac{K^2}{r^4}+6bC_0+
\frac{b^2}{4}(\frac{-B}{b-a}\frac{K^2}{r^3}+2C_0 r)^2}
+\frac{BK^2}{r^4}]e^{-b\widetilde{D(r)},
\end{eqnarray}
where $\widetilde{D(r)}=\frac{B}{2(b-a)}\frac{K^2}{r^2}+C_0 r^2$.

In Fig. 1, Fig. 2 and Fig. 3 we plot the $8\pi p$, $8\pi\rho$ and
$ [q(r)]^{2} $ against the parameter $r$ for some fixed values of
the integration constants $C_{0}$ and $C_{1}$ and parameters $A$,
$B$, $a$, $b$ and $K$.

\begin{figure}[ptb]
\begin{center}
\vspace{0.5cm}\includegraphics[width=0.6\textwidth]{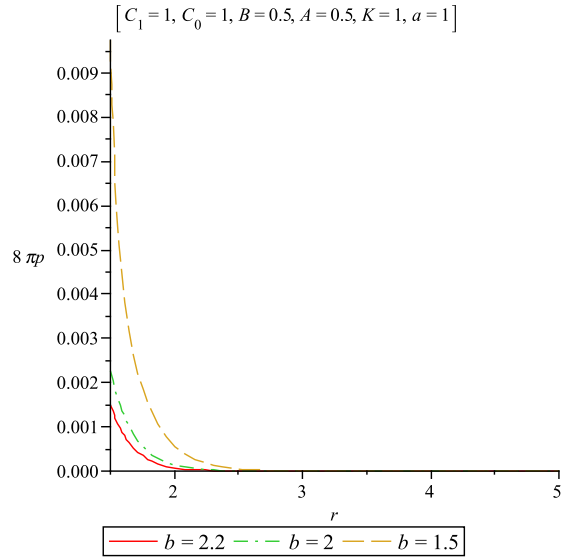}
\end{center}
\caption{Bayin's isotropic solutions with charge showing pressure
vs radius plot for the given specifications in the figure
(Case-4.1).} \label{Fig:1}
\end{figure}

\begin{figure}[ptb]
\begin{center}
\vspace{0.5cm}\includegraphics[width=0.6\textwidth]{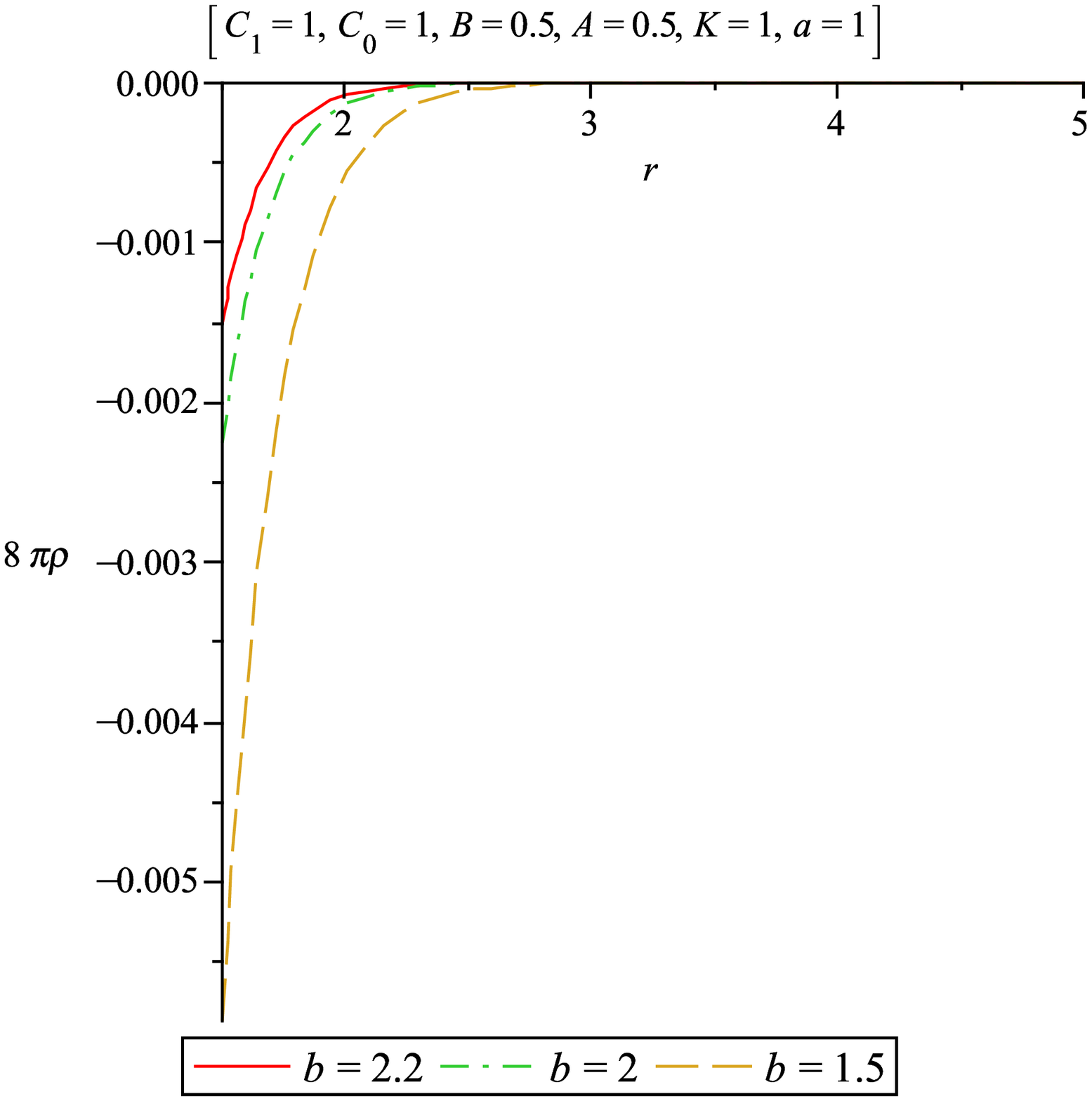}
\end{center}
\caption{Bayin's isotropic solutions with charge showing density
vs radius plot for the given specifications in the figure
(Case-4.1).} \label{Fig:2}
\end{figure}

\begin{figure}[ptb]
\begin{center}
\vspace{0.5cm}\includegraphics[width=0.6\textwidth]{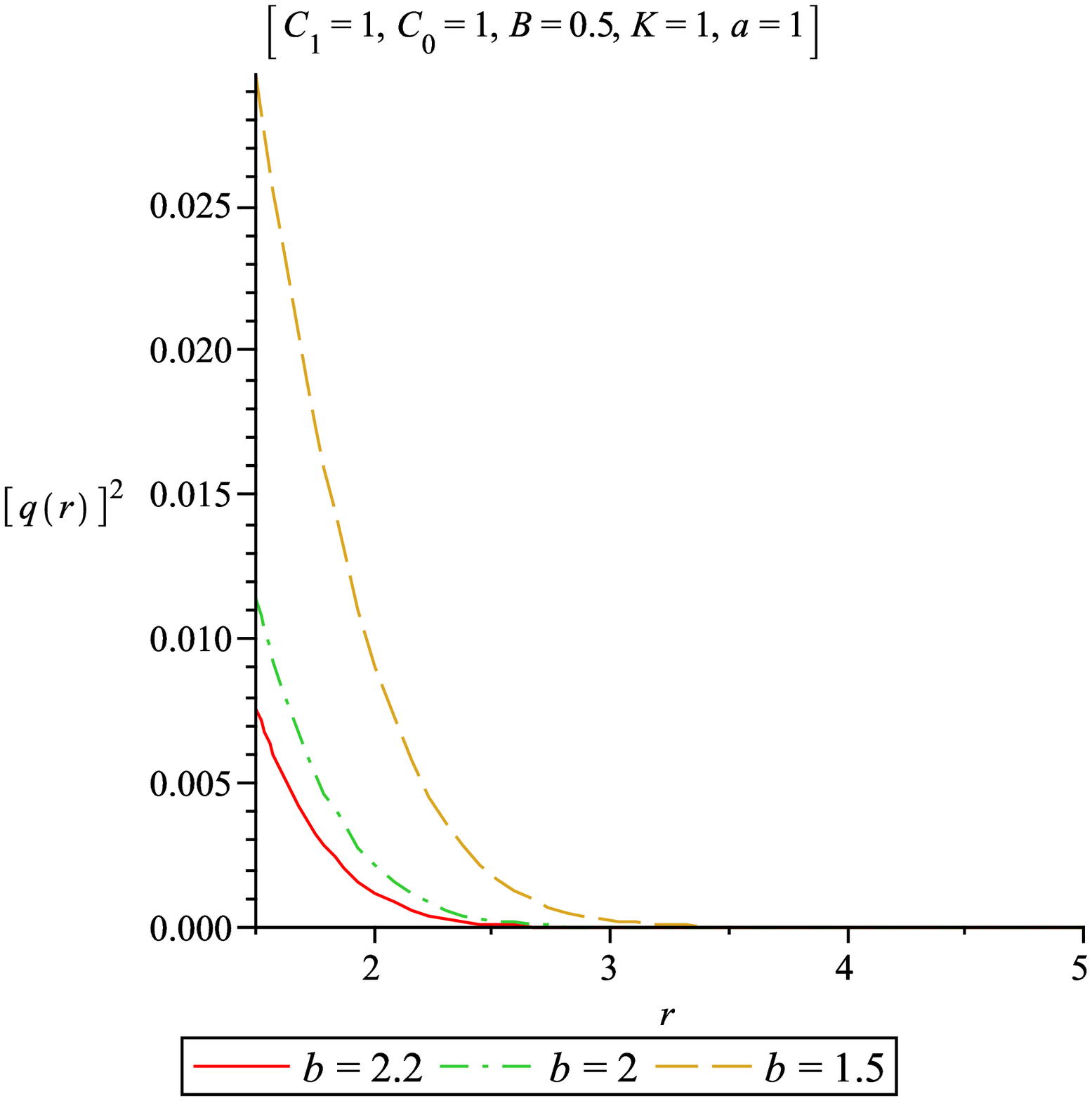}
\end{center}
\caption{Bayin's isotropic solutions showing charge vs radius plot
for the given specifications in the figure (Case-4.1).}
\label{Fig:3}
\end{figure}

\begin{figure}[ptb]
\begin{center}
\vspace{0.5cm}\includegraphics[width=0.6\textwidth]{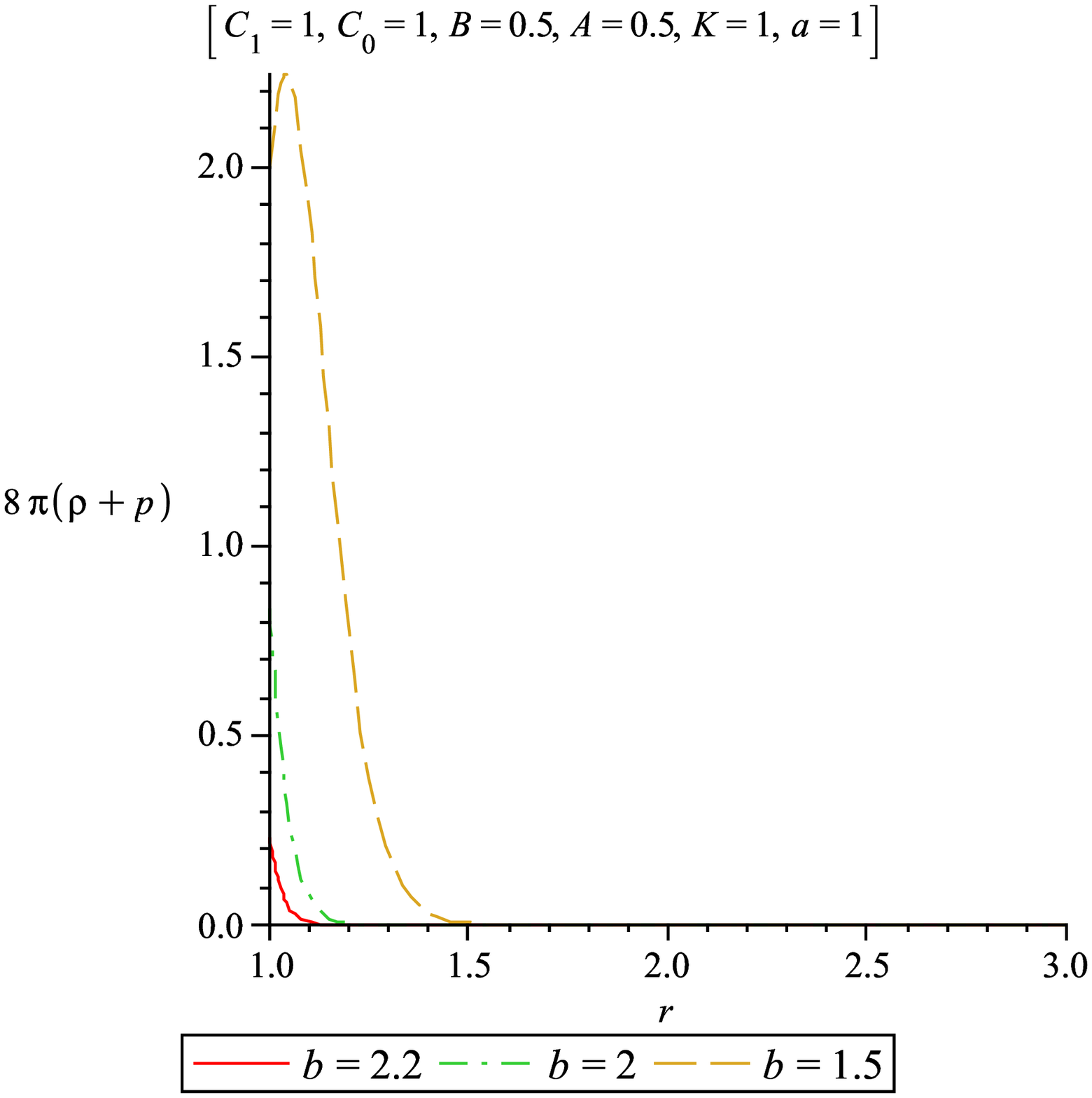}
\end{center}
\caption{Bayin's isotropic solutions with charge showing EOS
(pressure + density) vs radius plot for the given specifications
in the figure (Case-4.1).} \label{fig4}
\end{figure}

Fig. 4 above indicates that $P+\rho $ is positive. That means
though that the Weak Energy Condition (WEC) is violated in our
case but obeys the Null Energy Condition (NEC) as well as the
Strong Energy Condition (SEC) and the Dominant Energy Condition
(DEC). This is a very specific charged fluid solution. In
reference to the figure for p (Fig. 1), since $p=0$ gives the
radius of the charged fluid (rather we would say, $p(r=R)=0$), one
may conclude that the radius of the charged fluid falls within $3$
to $4$ Km. Thus our charged fluid is highly compact and we may
consider our charged fluid as typically a highly compact strange
star (we call it `strange' in the sense that EOS components follow
peculiar properties here i.e. they obey all energy conditions
except WEC). In this context we would like to mention that in the
original Bayin solutions, case V and case VI, $\rho $ is also
negative.

Therefore, following Bayin \cite{Bayin1978} we are now in the
position to find out pressure-to-density ratio at the centre of
the charged sphere in the form
\begin{eqnarray}
\frac{p_{c}}{\rho _{c}}=-\frac{b-2a}{b}.
\end{eqnarray}
~~~~We notice that the pressure-to-density ratio at the centre of
the charged sphere is entirely independent of the $A$ and $K$
parameters and integration constant $C_{1}$. In the case $b-2a>0$
with $b>0$ we have $\frac{p_{c}}{\rho _{c}}<0$ and this is not the
case with simultaneous positive pressure and density at the
origin. For $b-2a<0$ and $b>0$ we obtain $\frac{p_{c}}{\rho
_{c}}>0$, which is a physically meaningful result for positive
values for pressure and density at the origin. As a special case,
if we choose $a=0$, then we can easily recover the equation of
state related to the Cases I and II for $n=1$ of Ray et al.
\cite{Ray2007b} which refers to the `false vacuum' or `degenerate
vacuum' or `$\rho $-vacuum' equation of state
\cite{Davies1984,Blome1984,Hogan1984,Kaiser1984}. Due to the
repulsive nature of the pressure this provides a mechanism to
avert the problem of singularity at the centre. On the other hand,
if we choose $a=2$ and $b=3$ then we can recover the result of Ray
et al. \cite{Ray2007b} related to the Case I for $n=3$ which
refers to radiation. In Table 1 we have shown different
possibilities for different values of the parameters $a$ and $b$.

\begin{table}[tbp]
\caption{$\protect\rho_c$ and $p_c$ for different values of $a$
and $b$ related to the equation (21).} \label{tab1}
\begin{center}
\begin{tabular}{@{}llrrrrlrlr}
$Physical~status$ & $a$ & $b$ & $b-2a$ & $p_c/\rho_c$ &  &  &  & &
\\ \hline $Dust$ & $arbitrary$ & $2a$ & $0$ & $0$ &  &  &  &  &
\\ $Vacuum~fluid$ & $0$ & $arbitrary$ & $b$ & $-1$ &  &  &  &  &
\\ $Stiff~fluid$ & $1$ & $1$ & $-1$ & $+1$ &  &  &  &  &  \\
$Radiation$ & $2$ & $3$ & $-1$ & $1/3$ &  &  &  &  &  \\ \hline
\end{tabular}
\end{center}
\end{table}

Substituting $p(R)=0$, we can get an expression for the boundary
of the charged sphere as follows:\\

$b(b-2a)C_{0}^2R^{8}+2(b-a)C_{0}R^{6}-\frac{b(b-2a)BK^{2}C_{0}}{
(b-a)}R^{4}+\frac{b(b-2a)B^{2}K^{4}}{4(b-a)^{2}}R^{0} =0. $\\

Though it is a complicated power law equation for $R$, and hence
very difficult to solve yet from this we get the same expression
$R^{2}=-\frac{2(b-a)}{b(b-2a)C_{0}}$ which was obtained by Bayin
(1978) in his non-charge case (see equation (4.23) there in).

Again, comparing Reissner-Nordstr{\"{o}}m metric in isotropic
coordinates (16) [and hence (17) and (18)] we get
\begin{equation}
A{C_1}^{-a}e^{-a\widetilde{D(R)}}=\frac{\left(1-\frac{m^2}{4R^2}+\frac{q^2}{4R^2}\right)^2}{\left[\left(
1+\frac{m}{2R}\right)^2 -\frac{q^2}{4R^2}\right]^2},
\end{equation}
\begin{equation}
B{C_1}^{b}e^{b\widetilde{D(r)}}=\left[\left(1+\frac{m}{2R}\right)^2-\frac{q^2}{4R^2}\right]^2
\end{equation}

so that
\begin{equation}
C_0=\frac{-G\pm (G^{2}-4FH)^{1/2}}{2F}
\end{equation}
where
\begin{equation}
F=b(b-2a)R^8,
\end{equation}
\begin{equation}
G=2(b-a)R^6-\frac{b(b-2a)}{(b-a)}BK^2R^4,
\end{equation}
\begin{equation}
H=\frac{b(b-2a)}{4(b-a)^2}B^2K^4
\end{equation}
and
\begin{eqnarray}
C_1=\frac{\left(1-\frac{m^2}{4R^2}+\frac{q^2}{4R^2}\right)^{-2/a}}{\left[\left(
1+\frac{m}{2R}\right)^2 -\frac{q^2}{4R^2}\right]^{-2/a}}A^{1/a}e^{-\widetilde{D(R)}}.
\end{eqnarray}

Therefore, for $B$ we get
\begin{eqnarray}
B=\frac{\left[\left(
1+\frac{m}{2R}\right)^2 -\frac{q^2}{4R^2}\right]^{2(2a-b)/a}}
{\left(1-\frac{m^2}{4R^2}+\frac{q^2}{4R^2}\right)^{-2b/a}}A^{-b/a}.
\end{eqnarray}
~~~~ Therefore, considering arbitrary values for $a$, $b$ and $A$
we get expressions for $C_{0}$, $C_{1}$ and $B$ in terms of $m$
and $R$.

\subsection{The case for $c \neq1$ and $d=-\frac{1}{2}(b^{2}+a^{2})/(b-a)$}
\begin{eqnarray}
8 \pi p = f\left[\frac{b-2a}{4(1-c)}
\left(\frac{S^{\prime}}{S}\right)^2 +
\frac{b-a}{br}\left(\frac{S^{\prime}}{S}\right)\right] +
\frac{q^{2}}{r^{4}},
\end{eqnarray}
\begin{eqnarray}
8 \pi\rho=f \left[-\left(\frac{S^{{\prime}{\prime}}}{S }\right) +
\left(1-\frac{b}{4(1-c)}\right)\left(\frac{S^{\prime}}{S}\right)^2
- \frac{2}{r} \left(\frac{S^{\prime}}{S}\right)\right]
-\frac{q^{2}}{r^{4}}
\end{eqnarray}
where
\begin{equation}
f=\frac{bS^{-b/(1-c)}}{B(1-c)},
\end{equation}
\begin{equation}
S=\frac{B(1-c)}{2(b-a)}\frac{K^{2}}{r^{2}} + C_0 r^2 + C_1,
\end{equation}
\begin{equation}
S^{\prime}=-\frac{B(1-c)}{b-a} \frac{K^2}{r^3} + 2C_0 r,
\end{equation}
\begin{equation}
S^{{\prime}{\prime}}=\frac{3B(1-c)}{b-a} \frac{K^2}{r^4} + 2C_0.
\end{equation}

The pressure-to-density ratio at the centre of the charged sphere
in this case is as follows \cite{Bayin1978}:
\begin{eqnarray}
\frac{p_{c}}{\rho_{c}}= -1.
\end{eqnarray}

The same `false vacuum' or `degenerate vacuum' or `$\rho $-vacuum'
equation of state~\cite{Davies1984,Blome1984,Hogan1984,Kaiser1984}
is achieved again without imposing the condition $a=0$ in the
present case. Therefore, also due to the repulsive nature of the
pressure this provides a mechanism to avert the problem of
singularity at the centre.

The radius of the star is given by the expression obtained from
the condition $p(R)=0$. However, here $S$ and $S^{\prime }$ have
to be taken at $ r=R$, so that the final expression for the radius
$R$ can be written as
\begin{equation}
\frac{(b-2a)}{4(1-c)}\left(\frac{s^{\prime}}{s}\right)^2
R^4+\frac{b-a}{b}\left(\frac{s^{\prime}}{s}\right)
R^3+\frac{1-c}{b}q^2B{s^{b/(1-c)}}=0
\end{equation}
where we have used the symbol $s=S(R)$.

We can find out the values of $C_{0}$, $C_{1}$ and $B$ in terms of
$M$ and $ R $ from the equation (37) and the following boundary
conditions
\begin{eqnarray}
As^{-a/(1-c)} =\frac{\left(1-\frac{m^2}{4R^2}+\frac{q^2}{4R^2}\right)^2}{\left[\left(
1+\frac{m}{2R}\right)^2 -\frac{q^2}{4R^2}\right]^2},
\end{eqnarray}
\begin{eqnarray}
Bs^{b/(1-c)} =\left[\left(1+\frac{m}{2R}\right)^2-\frac{q^2}{4R^2}\right]^2.
\end{eqnarray}

Due to the complicated nature of the solutions above, we could not
discuss these analytically, especially we are unable to find out
the polytropic relation as obtained by Bayin \cite{Bayin1978} in
his equation (4.43) for the classical limit $P_c/\rho
_c\rightarrow 0$ \cite{Bludman1973}. We expect to provide this
study in a future project.\\

\section{Conclusions}
~~~~The study of static charged fluid spheres in general
relativity is far from being complete, but some recent studies
have motivated us to extend one previous work~\cite{Ray2007b} for
improving the understanding of this topic. We found out the
solutions of field equations in isotropic cases of charged fluid
spheres in general relativity and performed a detailed study of
the analytical solutions. In order to obtain the analytical
solutions of the Einstein-Maxwell field equations, we made use of
some assumptions and introduced two specialization, the first with
the specification $c=1$ and $d=-b$ and the second with the values
$c\neq 1$ and $d=-\frac{1}{2} (b^{2}+a^{2})/(b-a)$. We calculated
the corresponding expressions for the function $\phi (r)$ and for
the metric coefficients $e^{\nu }$ and $e^{\mu }$, which depend on
the parameters $A$, $B$, $a$, $b$ and integration constants
$C_{0}$ and $C_{1}$. Matching the interior solution to the
exterior Reissner-Nordstr\"{o}m metric in isotropic coordinates at
$r=R$ we obtained the formulae for $e^{\nu (R)}$ and $e^{\mu
(R)}$.

We performed a deeper investigation of the solutions and for the
case $c=1$ and $d=-b$ we established the expression for the
pressure-to-density ratio at the origin given by
$\frac{p_{c}}{\rho _{c}}=-\frac{b-2a}{b}$ and the imposed
conditions for $a$ and $b$ for obtaining a physically meaningful
model for the star. For a zero value of $a$ we obtained the case
of the \textit{false vacuum} or \textit{degenerate vacuum} or
\textit{$\rho $-vacuum}
\cite{Davies1984,Blome1984,Hogan1984,Kaiser1984}. In addition, we
found out the explicit expression for the radius $R$ and discussed
the physically reasonable cases. The radius $R$ for the charged
sphere has a completed expression which, however, reduces to the
same expression as in the non-charge case of Bayin~\cite{Bayin1978}. 
Finally, comparing Reissner-Nordstr\"{o}m metric
in isotropic coordinates we established the expressions for the
integration constants $C_{0}$ and $C_{1}$ and $B$ parameter in
terms of mass $m$ and the radius of the star $R$. On the other
hand for the specification $c\neq 1$ and
$d=-\frac{1}{2}(b^{2}+a^{2})/(b-a)$ we computed the pressure $p$
and the density $\rho $ that present a dependence on a new
function $S(r)$ and its first and second derivatives with respect
to the $r$ coordinate. In this case the expression for the
pressure-to-density ratio at the origin can directly be given by
$\frac{p_{c}}{\rho _{c}}=-1$. We also gave the expressions that
allow to evaluate the boundary $R$ of the star and the values of
$C_{0}$, $C_{1}$ and $B$ in terms of $m$ and $R$, but due to their
complicated form we plan to perform a detailed study in a future
work.

The most straight forward observation in the present work is that
from our results we are able to recover the neutral cases of Bayin
\cite{Bayin1978}. We also observe that the models presented here
have negative energy density and positive pressure and support the
conclusion that static charged fluid spheres are connected to
compact strange stars. We conclude that in the last years there is
progress in the study of static charged fluid spheres which are
connected to compact stars, but deeper and wider investigations
for searching for new solutions need to be performed. One
possibility will be to investigate our scenario based on different
specialization and obtain new analytical solutions that can lead
to other particular cases and more constrained connections with
the involved parameters.\\

\section*{Acknowledgements}
FR and SR are thankful to the authority of Inter-University Centre
for Astronomy and Astrophysics, Pune, India for providing research
facilities.

\end{document}